\documentstyle[epsfig,12pt]{article}
\newcommand{\bce}{\begin{center}}
\newcommand{\ece}{\end{center}}
\newcommand{\beq}{\begin{equation}}
\newcommand{\eeq}{\end{equation}}
\newcommand{\bea}{\vspace{0.25cm}\begin{eqnarray}}
\newcommand{\eea}{\end{eqnarray}}

\newcommand{\bsigma}{\mbox{\boldmath $\sigma$}}

\newcommand{\ba}{\begin{array}}
\newcommand{\ea}{\end{array}}

\def\lsim{\mathrel{\rlap{\lower4pt\hbox{\hskip1pt$\sim$}}
    \raise1pt\hbox{$<$}}}         
\def\gsim{\mathrel{\rlap{\lower4pt\hbox{\hskip1pt$\sim$}}
    \raise1pt\hbox{$>$}}}         

\def\lsim{\mathrel{\rlap{\lower4pt\hbox{\hskip1pt$\sim$}}
    \raise1pt\hbox{$<$}}}         
\def\gsim{\mathrel{\rlap{\lower4pt\hbox{\hskip1pt$\sim$}}
    \raise1pt\hbox{$>$}}}         

  \advance\oddsidemargin  by -1in
  \advance\evensidemargin by -1in
  \advance\topmargin      by -0.5in
\def\lsim{\mathrel{\rlap{\lower4pt\hbox{\hskip1pt$\sim$}}
    \raise1pt\hbox{$<$}}}         
\def\gsim{\mathrel{\rlap{\lower4pt\hbox{\hskip1pt$\sim$}}
    \raise1pt\hbox{$>$}}}         

\def\beq{\begin{equation}}
\def\endeq{\end{equation}}
\def\arr{\begin{eqnarray}}
\def\endarr{\end{eqnarray}}
\makeindex
 
\begin{document}

\begin{flushright}
 ITEP-PH-1/2004
\end{flushright}
\vspace{1cm}

\begin{center}
{\Large \bf
Absorption in Ultra-Peripheral 
Nucleus-Atom Collisions in Crystal 

\vspace{1.0cm}}
 
{\large \bf
V.R. Zoller\medskip\\ }
{\sl  Institute for Theoretical and Experimental Physics \\
117218 Moscow Russia
{\footnote {\rm zoller@itep.ru}}\vspace{1cm}\\}
{\bf           Abstract}
\end{center}

 The Glauber theory description of 
 particle- and nucleus-crystal Coulomb interactions at high-energy 
is developed. The allowance for the lattice thermal vibrations  
is shown to produce  strong  absorption effect which is of prime importance for
quantitative understanding of  the coherent Coulomb
excitation of ultra-relativistic particles and nuclei passing through the 
crystal.

\newpage

In this communication we discuss the origin and estimate 
the strength of the  absorption effect in 
coherent  particle- and nucleus-crystal Coulomb interactions at high-energy.

Generally, the  multi-loop corrections  generate the 
imaginary part of the scattering amplitude  even  if 
the tree-level amplitude is  purely real. For example, 
the purely real Born amplitude of the 
  high-energy Coulomb scattering in crystal
  acquires the imaginary part due to the multiple scattering (MS) effects.
 However, in the widely used
 static/frozen  lattice  
approximation (SL approximation)
 the account of rescatterings  alters only the overall real phase of 
the full amplitude  thus producing no absorption effect. 
The latter is related to the 
creation and annihilation of excited intermediate 
 states of crystal and as such
manifests itself only beyond the SL approximation.  
With   the allowance for the 
lattice thermal vibrations the Coulomb phase shift function gets
 non-vanishing imaginary part which is interpreted as an  absorption 
effect. The account of the  lattice thermal vibrations provides
 a natural ultra-violet regulator of the theory
and, as we shall see, enables  
quantitative understanding of the phenomenon of the coherent Coulomb
excitation of relativistic particles and  nuclei passing through the crystal.
The latter is the goal we pursue in  this communication.

We start with 
the well known example of  the coherent Coulomb  elastic scattering 
of charged  particle (charge $Z_1$) by  a linear chain of $N$ identical atoms  
in a crystal target.
The interatomic distances in crystal, $a$,  are large,  
compared to the  
 Thomas-Fermi screening radius $r_{0}$, 
$a\sim 3-5\AA\gg r_{0}= r_B Z_2^{-1/3}\sim 0.1 \AA$,
 where $Z_2$ is the atomic number of the target atom and $r_B$
 is the Bohr radius \cite{GEM}. 
The relevant impact parameters, $b$, 
satisfy the condition  $b\ll a$ and  the amplitudes of 
scattering by different atomic chains  parallel to a given  crystallographic 
axis are incoherent.

The   amplitude of small-angle elastic  scattering in  the eikonal 
approximation \cite{GLAUBER}  can be represented in the following form
\bea
F(q)=
{ip}\int bdb J_0(qb)\left\{1-\langle \exp[i\chi({\mu b})]\rangle^N\right\}.
\label{eq:EL5} 
\eea
where  ${\bf q}$ is the $2D$-vector of the momentum transfer and
the incident particle momentum $p$ is assumed  to be large enough to satisfy
the condition of applicability of the straight paths approximation,
$p/q^2\gg aN$. The latter condition insures the coherence of 
interactions with different atoms. 
In eq.(\ref{eq:EL5})
the screened Coulomb phase shift function is  
\beq
\chi(\mu b)  =-\beta K_0(\mu b),
 \label{eq:CHI} 
\endeq
with $\beta=2\alpha Z_1Z_2$ and $\mu=r_0^{-1}$.
Hereafter,  $J_{0,1}(x)$ and $K_{0,1}(x)$ are the  Bessel functions.
 The brackets $\langle\,\,\,\rangle$ signify that an 
average is to be taken over all configurations of atoms in 
the ground state of crystal. We neglect all position correlations of the atoms 
and describe the ground state of crystal
by the   wave function $|\Psi\rangle\propto 
\prod_{j=1}^N |\psi_j\rangle$.

After the azimuthal integration  the term
$\langle\exp(i\chi)\rangle$ takes the form
\bea
\langle\exp(i\chi)\rangle
=\int d^2{\bf s}\rho({s})
\exp[i\chi(\mu |{\bf b}-{\bf s}|)]\nonumber\\
=\exp(-{\Omega^2b^2})\int dx
\exp(-x)\nonumber\\ 
\times I_0({2b\Omega\sqrt{x}})
\exp\left[-i\beta K_0(\mu\sqrt{x}/\Omega)\right]. 
\label{eq:EL8} 
\eea
The $2D$-vector ${\bf s}$, describes the
position of the target atom 
in the impact parameter plane.   
The one-particle probability distribution  $\rho({s})$   is as follows
\beq
\rho({s})=\int dz |\psi({\bf s},z)|^2=
(\Omega^2/\pi)\exp\left(-\Omega^2{\bf s}^2\right). 
\label{eq:RHO}
\endeq
For the most commonly studied elements at room temperature the ratio
$\mu/\Omega$  varies in a wide  range, from $\mu/\Omega\sim 0.1$  
to  $\mu/\Omega\sim 1$ \cite{GEM}. 
Consider first the region of  small impact parameters. 
For  $ b\ll 1/2\Omega$ only small $s$,
such that $\mu s\lsim 1$,  contribute.   One can put  then
in eq.(\ref{eq:EL8}) $K_0(\mu s)\simeq \log(1/\mu s)$ 
and integrate over $s$. The result is
\bea
\langle \exp(i\chi)\rangle
=\left({\mu\over\Omega}\right)^{i\beta}
\Gamma\left(1+{i\beta\over 2}\right)
\!\Phi \left(-{i\beta\over 2};1;-\Omega^2b^2\right).
\label{eq:HB6} 
\eea
In eq.(\ref{eq:HB6})
$\Phi(a,b;z)$
is the confluent hyper-geometric function. 
From (\ref{eq:HB6}) it follows, in particular,  that 
\bea
\left|\langle \exp(i\chi) \rangle\right|_{b=0}=
\left[{\pi \beta\over 2\sinh(\pi\beta/2)}\right]^{1/2}.
\label{eq:EL10} 
\eea
and in the weak coupling regime,  $\beta\ll 1$,
\bea
\left|\langle \exp(i\chi) \rangle\right|_{b=0}
\simeq 1- {\pi^2\beta^2\over 48},
\label{eq:HB79}
\eea 
while for $\beta\gsim 1$
\bea
\left|\langle \exp(i\chi) \rangle\right|_{b=0}
\simeq \sqrt{\pi\beta}\exp\left(-{\pi\beta \over 4 }\right).
\label{eq:HB9}
\eea
Therefore, at small impact parameters, $b\lsim \Omega^{-1}$,
the intensity of outgoing nuclear waves as a function of $N$
 exhibits the exponential attenuation.

 The absorption effect becomes weaker toward the  
region of large impact parameters $b\gsim 1/2\Omega$,
\bea
\left|\langle \exp(i\chi) \rangle\right|^N\simeq 
\left|\langle \exp(i\chi) \rangle\right|^N_{b=0}\nonumber\\
\times\left[1+{N\beta^2\over 16 }(\Omega b)^4+...\right].
\label{eq:HB10}
\eea  
For still larger $b$,  $b\gg 1/2\Omega$, making use of the
asymptotic form $I_{0}(z)\simeq (2\pi z)^{-1/2}\exp(z)$ and
the condition
\bea
\omega={d\chi\over db}=\mu\beta K_1(\mu b)\ll \Omega.
\label{eq:omega}
\eea 
yields
\bea
\langle\exp(i\chi)\rangle
\simeq 2\Omega\int {sds\over \sqrt{\pi bs}}\exp[-\Omega^2(b-s)^2]
 \exp[i\chi(\mu s)]\nonumber\\
\simeq
 \exp(i\chi)\exp[-\omega^2/4\Omega^2].
\label{eq:HB2}
\eea
From (\ref{eq:HB2},\ref{eq:HB6},\ref{eq:EL5})  it follows that the 
absorption is especially strong  for  impact parameters
\bea
b\lsim b_{a}= {1\over 2\mu}\log{\pi\mu^2\beta^2 N \over 4\Omega^2}.
\label{eq:RBD} 
\eea
For  $b\lsim  b_a$ the atomic chain  acts like an opaque ``black'' disc.
Certainly, the value of this finding  differs  for different observables 
and for different  processes proceeding
 at different impact parameters.

Integrating once  by parts reduces $F(q)$ to the form convenient
for evaluation of the total cross section,
\bea
F(q)={ip\mu N\over q}\int_0^{\infty} bdbJ_1(qb)
\langle i\chi^{\prime}\exp(i\chi)\rangle
\langle\exp(i\chi)\rangle^{(N-1)}.
\label{eq:HB8} 
\eea
At small $q$ and large $N$ only large impact parameters,
 $b\gg \mu^{-1}$, may contribute to $F(q)$.
This is the multiple scattering effect \cite{GLAUBER,KALASH} which gives 
rise to the dominance 
of ultra-peripheral collisions in the coherent particle-crystal interactions.
Then, for   $q\lsim q_0= \mu/\xi$
 and $\xi\gg 1$, the steepest descent
from the saddle-point
\bea
b_0=\mu^{-1}[\xi+i\pi/2]
\label{eq:B}
\eea
in eq.(\ref{eq:HB8}) yields
\bea
F(q)\simeq {ipb_0\over q}J_1(qb_0).
\label{eq:FQ0}
\eea
The effect of lattice thermal vibrations at small $q$ 
appears to be  marginal and reduces to the factor
$\exp({\mu^2/ 4\Omega^2 N})$ in (\ref{eq:FQ0})
which is irrelevant in the region of  large $N$ where
the amplitude $F(q)$ 
coincides with the elastic scattering amplitude given by the SL 
approximation \cite{KALASH}.

If  $q\gsim q_0$ the stationary phase approximation gives the elastic
 scattering amplitude of the form
\bea
F(q)\simeq {-ip\sqrt{\eta}\over \mu q }
\exp\left(-{iq\eta\over \mu}\right)
\exp\left(-{q^2\over 4\Omega^2 N}\right)
\label{eq:FQQ}
\eea
 where $\eta=\log(\mu\beta N/q) \gg 1$. 
This is the   lattice  vibrations which provide a natural momentum cutoff and  
insure the convergence of the integral  for the coherent elastic scattering 
cross section,
\bea
\sigma_{el}={\pi\over p^2} \int {dq^2}|F(q)|^2
\approx {\pi \xi^2\over \mu^2}\int_0^{q_0^2}{dq^2\over q^2}
J^2_1\left({q\xi\over\mu}\right)\nonumber\\
+ {\pi\over \mu^2}\int_{q_0^2}^{\infty}
{dq^2\over q^2}\log\left({\mu\beta N\over q}\right)\exp
\left(-{q^2\over 2\Omega^2 N}\right),
\label{eq:SIGEL}
\eea 
which for $\xi\gg 1$ is simply
\bea
\sigma_{el}
\approx {\pi \over \mu^2}\xi^2.
\label{eq:SIGEL1}
\eea 
From eq.(\ref{eq:FQ0}) by means of the optical theorem we find 
the total cross section 
\bea
\sigma_{tot}={4\pi\over p}Im F(0)
\approx {2\pi\over\mu^{2}} \xi^2. 
\label{eq:STOT}
 \eea
Consequently,
at high-energy  and  for   $\xi\gg 1$,
$ \sigma_{el}\approx {1\over 2}\sigma_{tot}$.

Now let us turn to the process of the 
 coherent Coulomb excitation of ultra-relativistic
particles and  nuclei passing through the  crystal.
This way of the experimental study of rare processes  has been proposed 
in \cite{PROP1, PROP2,PIVOVAR,PROP3,PROP4,FUSINA,DUBIN}.

 The  ultra-relativistic
projectile-nucleus 
(the mass number $A$, the  charge $Z_1$ and  the four-momentum $p$) 
 moving along a crystal  axis undergoes a correlated series of 
 soft  collisions
which give rise to diagonal ($A\to A$, $A^*\to A^*$) and
 off-diagonal ($A\to A^*$, $A^*\to A$)
 transitions. 
  In \cite{FUSINA,PROP1,PROP2} it has been proposed to study
the electric dipole transition in $^{19}F$,
  the excitation of the state $|J^{\pi}={1/2}^-\rangle$ 
from the ground state $|1/2^+\rangle$. 
The phenomenological   matrix element of the transition $1/2^+\to 1/2^-$
is \cite{NUCSTAT}
\beq
{\cal M}={1\over 2}d\bar u(p^{\prime}) 
\gamma_5 \left(\hat q\hat\varepsilon- 
\hat\varepsilon\hat q\right) u(p), 
\label{eq:M}
\endeq
where both $u(p^{\prime})$ and $ u(p)$ are bispinors of initial
 and final states of the
projectile,
 $d$ is the transition
 dipole moment and  $\varepsilon$
 is the photon polarization vector. 
The  transverse and longitudinal components the 4-vector 
$p-p^{\prime}$ are  denoted by  ${\bf q}$ and $\kappa$, respectively.
In what follows $q=|{\bf q}|$. 
Because of large value of  the life-time  
of the $110$ KeV   level $^{19}F(1/2^-)$ \cite{AJZEN}
the decay of excited state inside the target crystal can be safely neglected.  
Due to the smallness of  the transition dipole moment, 
$d\simeq 5\times 10^{-8}$ KeV$^{-1}$,  the excitation amplitude
is much smaller than the elastic Coulomb 
amplitude for all $q$ up to $q\sim \sqrt{4\pi\alpha}Z_1/d$ and can
 be considered as a perturbation \cite{NUCSTAT}. 
Thus, the multi-channel problem
reduces to the one-channel one.

The high-energy helicity-flip Born amplitude 
 of the transition $1/2^+\to 1/2^-$ in  collision of the 
projectile-nucleus with $N$  bound atoms in crystal reads 
\beq
{F}^{B}_{ex}({\bf q})=S(\kappa){p\over 2\pi}
{g(\bsigma{\bf q})
\over q^2 +\lambda^2 }\exp\left(-{q^2\over 4\Omega^2}\right)\,,               
\label{eq:TB}
\endeq 
where  $\bsigma=(\sigma_1,\sigma_2,\sigma_3)$ 
is the Pauli spin vector, 
$\{\sigma_i,\sigma_j\}=2\delta_{ij}$ and the amplitude we are constructed
 is to be regarded as an operator 
which transforms the initial helicity state of the projectile into its final 
state. 
In the denominator of eq.(\ref{eq:TB})
 $\lambda^2=\mu^2+\kappa^2$.
In the Glauber approximation the longitudinal momentum transfer
which determines the coherency length, $l_c\sim \kappa^{-1}$,
 reads  \cite{GRIBOV}
\beq
\kappa={M\Delta E\over p},
\label{eq:kappa}
\eeq
where $M$ is  the mass of  
 projectile and   $\Delta E$ is the excitation energy 
\footnote{The Fresnel corrections
to geometrical optics which are neglected here  become important
for large $N$ or  large $q$ diminishing the coherency length 
and  bringing about an additional suppression of coherent processes
\cite{NEUT}.}.

 The  structure factor of crystal 
$S(\kappa)$ to the first order in $g$   is  
\beq
S(\kappa)=\exp\left[-{\kappa^2\over 4\Omega^2} \right]
{{\sin(\kappa Na/2)}\over{\sin(\kappa a/2)}}\,.
 \label{eq:SLNEW}
\endeq 
If the projectile momentum  satisfies 
 the resonance condition \cite{PROP1,PROP2,FUSINA,PROP3} 
\beq
 {M\Delta E\over p}={2\pi n\over a}\,,\,\,\,n=0,\,1,\, 2...\label{eq:KAPRES}
\endeq
$S(\kappa)\sim N$.
 Then, to the first order in $g$
(Born approximation)
  the cross section of the coherent
excitation of the projectile in scattering on a chain of $N$ atoms in 
crystal is     
\bea 
\sigma^B_{ex}={\pi\over p^2}\int dq^2 |F^B_{ex}({\bf q})|^2
\propto {g^2N^2\over 4\pi}
\label{eq:SIGBORN}
\eea
where 
$g=\sqrt{4\pi\alpha}dZ_2$.
The central idea of \cite{PROP1, PROP2,PIVOVAR,PROP3,FUSINA,DUBIN} 
based on the Born approximation is that the  transition rate can be 
enhanced substantially due to coherency
of interactions which is assumed  to sustain  over the large distance scale. 
 The law  
$\sigma_{ex}\propto\,N^2$
is expected to hold true  up to the crystal thicknesses    
$N=L/a\sim 10^5-10^6 $ in tungsten target.
In \cite{DUBIN} the Born approximation for the coherent excitation
of $\Sigma^+$ in high-energy  proton-crystal interactions
$p\gamma\to \Sigma^+ $  has been assumed 
to be valid up to $N\sim 10^8$. However,  the account of the initial and final
state  Coulomb interactions  dramatically 
changes the dependence of $\sigma_{ex}$ on $N$.
For example, even in the diamond crystal   
\bea
\sigma_{ex}\sim \sigma^B_{ex}
\left(1-{N\omega^2\over 2\Omega^2}\right)
\label{eq:SBCOR}
\eea
and $\omega^2/2\Omega^2\simeq 2\beta^2\mu^2/\Omega^2\sim 1/20$ (see 
\cite{JETP} for more details).
Thus,
the Born approximation becomes irrelevant  already  at $N\gsim  10$.

 The evaluation of the   transition amplitude on a chain 
 of $N$ identical atoms  including all the multi-photon
 t-channel exchanges reads  
\beq
F_{ex}({\bf q })={p\over \pi}
\int {d^2{\bf b}}\exp(i{\bf q} {\bf b})
\langle {f}^B_{ex}
\exp(i\chi)\rangle
\langle \exp(i\chi)\rangle^{N-1}
\label{eq:TFULL} 
\endeq
The  eq.(\ref{eq:TFULL}) contains two bracketed factors.  
The first one corresponds to the nuclear excitation amplitude
in scattering on  the atom bound in crystal.
At small impact parameters, $b\lsim 1/2\Omega $,
\bea
\langle {f}_{ex}^{B} \exp(i\chi)\rangle\simeq
S(\kappa){g\over 2\pi b}{(\bsigma {\bf n_b})}
\sinh\left({1\over 2}\Omega^2b^2\right)\exp\left(-{1\over 2}\Omega^2b^2\right).
\label{eq:GB22}
\eea
Because of both the multiple scattering effect and  absorption   only 
 $b\gg \mu^{-1}$ may contribute to $F_{ex}({\bf q })$. 
In this region of impact parameters
\bea
\langle {f}_{ex}^B \exp(i\chi)\rangle
\simeq
S(\kappa){g\over 4\pi}(\bsigma {\bf n_b})\lambda K_1(\lambda b)
\exp(i\chi)\exp\left(-{\omega^2\over 4\Omega^2}\right).
\label{eq:GB2}
\eea
The second factor in (\ref{eq:TFULL}) describes the 
initial and final state interactions  of the projectile   
and has been calculated above. 
Then,
\bea
F_{ex}({\bf q })\approx {gp\over 2\pi}S(\kappa)
(\bsigma{\bf n_q})\int_{1/\mu}^{\infty} 
bdbJ_1(qb)\nonumber\\
\times
\lambda K_1(\lambda b)\exp(iN\chi)\exp(-N\omega^2/4\Omega^2), 
\label{eq:TREDUCE} 
\eea
where  ${\bf n_q}={\bf q}/|{\bf q}|$. 
The contribution of the domain
$q\lsim q_0=\mu/\xi$  to 
the excitation  cross section 
can be neglected as far as  $F_{ex}\propto q$ for $q\lsim q_0$ . 
If $q\gg q_0$ and $\xi\gg 1$,
 the stationary phase approximation gives the coherent excitation
 amplitude of the form
\bea
F_{ex}({\bf q })\approx 
{ipg(\bsigma{\bf n_q})\over 2\pi \beta}{S(\kappa)\over N}
{\lambda\over\mu}{\sqrt{\eta}}
\exp\left(-\delta\eta\right)\nonumber\\
\times\exp\left(-{iq\eta\over \mu}\right)
\exp\left(-{q^2\over 4\Omega^2 N}\right).
\label{eq:TLARGE}
\eea
We see that the helicity-flip dynamics removes the factor $1/q$
from the elastic  amplitude (\ref{eq:FQQ}) thus making the 
UV-regularization
of the excitation cross section indispensable.
The latter is evaluated as,
\bea
\sigma_{ex}={\pi\over p^2}\int dq^2|F_{ex}({\bf q })|^2\nonumber\\ 
\sim  {g^2N^{1-\delta}\over 8\pi}C
\log\left({N\over\delta\gamma}\right),
\label{eq:SIGEX}
\eea
where $C=\gamma^{\Delta}\Delta^2\Gamma(\Delta)$, 
$\gamma=2\Omega^2/\beta^2\mu^2$, $\Delta=\lambda/\mu$ 
and $\delta =\Delta - 1\sim \kappa^2/2\mu^2\ll 1$. In (\ref{eq:SIGEX})
we put simply $S(\kappa)=N$.
Thus, the account of  multiple scatterings and absorption turns
the Born approximation cross section   
$\sigma_{ex}\propto N^2$ into $\sigma_{ex}\propto N^{1-\delta}\log N$.
 In the limit of $p\to \infty$ and $\delta\to 0$,
\bea
\sigma_{ex}
\sim  {g^2N\over 8\pi}\gamma
\log\left({N\over \gamma}\right)
\label{eq:SIGEX0}
\eea
The dependence of $\sigma_{ex}$ on $N$ differs from that of fully
unitarized $\sigma_{el}\propto \log^2 N$. The reason is that
  in $\sigma_{ex}$ we sum 
the eikonal diagrams to all orders in $\beta$ but only to the first order
in $g$. 
Such a  unitarization procedure is, of course, incomplete, but 
 this is of no importance for practical purposes since
the smallness of $d^2\Omega^2$ makes the next to leading order corrections 
negligibly small up to $N\sim \alpha Z_1^2/\delta\Omega^2d^2\sim 10^{12}$.

{\bf Acknowledgments:}   Thanks are due to N.N. Nikolaev for
useful comments.

\end{document}